\documentclass[conference]{IEEEtran}
\usepackage{ifpdf}
\usepackage{cite}
\ifCLASSINFOpdf
 \usepackage[pdftex]{graphicx}
\else
 \usepackage[dvips]{graphicx}
\fi
\usepackage{amsmath}
\usepackage{amsfonts}

\usepackage[ruled,vlined]{algorithm2e}

\usepackage{array}

\ifCLASSOPTIONcompsoc
   \usepackage[caption=false,font=normalsize,labelfont=sf,textfont=sf]{subfig}
\else
  \usepackage[caption=false,font=footnotesize]{subfig}
\fi
\usepackage{fixltx2e}
\usepackage{stfloats}
\usepackage{url}
\usepackage{xcolor}

\newcommand{\SubItem}[1]{
    {\setlength\itemindent{15pt} \item[-] #1}
}
\usepackage{xcolor}
\usepackage{xspace}
\newcommand\eat[1]{}

\newcommand{\mynote}[3]{%
  \ifthenelse{\boolean{showcomments}}{%
   \fbox{\bfseries\sffamily\scriptsize#1}%
   {\small$\blacktriangleright$\textsf{\emph{\color{#3}{#2}}}$\blacktriangleleft$}}%
  {%
   \@bsphack
   \@esphack
  }%
}

\newcommand{\edit}[1]{\textcolor{black}{#1}}
\newcommand{\name}{{X-Poly}\xspace}

\begin{document}
%
\title{Accelerating Polynomial Modular Multiplication with Crossbar-Based Compute-in-Memory}


\author{\IEEEauthorblockN{Mengyuan Li*, Haoran Geng*, Michael Niemier,
Xiaobo Sharon Hu}
\IEEEauthorblockA{Department of Computer Science and Engineering\\
University of Notre Dame, Notre Dame, IN, USA, 46556 \\
E-mail: mli22@nd.edu  hgeng@nd.edu mniemier@nd.edu shu@nd.edu \\ * These authors contributed equally. } }

\maketitle
\pagestyle{plain}


%


\pdfoutput=1
\begin{abstract}
Lattice-based cryptographic algorithms built on ring learning with error theory are gaining importance due to their potential for providing post-quantum security. However, these algorithms involve complex polynomial operations, such as polynomial modular multiplication (PMM), which is the most time-consuming part of these algorithms.  Accelerating PMM is crucial to make lattice-based cryptographic algorithms widely adopted by more applications. This work introduces a novel high-throughput and compact PMM accelerator, {\em \name}, based on the crossbar (XB)-type compute-in-memory (CIM).  We identify the most appropriate PMM algorithm for XB-CIM. We then propose a novel bit-mapping technique to reduce the area and energy of the XB-CIM fabric, and conduct processing engine (PE)-level optimization to increase memory utilization and support different problem sizes with a fixed number of XB arrays. \name design achieves $3.1 \times 10^{6}$ PMM operations/s throughput and offers 200$\times$ latency improvement compared to the CPU-based implementation. It also achieves 3.9$\times$ throughput per area improvement compared with the state-of-the-art CIM accelerators.

 \end{abstract}

\section{Introduction}
\label{sec:introduction}

\eat{
Overview of lattice-based cryptographic algorithms (HE and others) and their importance.
\begin{itemize}
    \item  Polynomial modular multiplication and its role in these algorithms: 
    \SubItem{ Ring learning with error is the key of this type of algorithm and involves lots of polynomial operations.}
    \SubItem{Polynomial modular multiplication is the most time-consuming part in these algorithms. Example: HE time breakdown.}
   \item Discuss existing research on accelerating polynomial modular multiplication and the limitations of current solutions.
   \SubItem{Most of them are NTT-based solutions, including ASIC, FPGA, and Compute-in-Memory(CIM) solutions.}
   \SubItem{Especially, CIM-based solution gains lots of benefits due to their speed.}
   \SubItem{However, If NTT-based solutions are best implementations in CIM paradigm?  We evaluate the suitability of NTT-based solutions for CIM paradigm (especially XBAR-type)}
    \item This work’s contribution:
    \SubItem {Design a PMM CIM accelerator (non-NTT type) and first evaluate the suitability of NTT-based solutions for CIM paradigm (especially XBAR-type and discuss the pros and cons of conventional conv1d and NTT-type PMM in CIM.}
    \SubItem {Propose a new xbar data mapping for high-bitwidth data and reduce almost a bunch of find-grained shift-add operations.}
    \SubItem {PE-level optimization to increase memory utilization and support the different scales of problems with a fixed number of xbar.}
\end{itemize}
}

Post-quantum cryptography (PQC) represents a critical area of research in the field of cryptography, driven by the impending threat posed by quantum computing to current cryptographic systems~\cite{PQR}. Many of these cryptosystems are currently being considered potential PQC candidates to address the challenges posed by quantum computing. Among these cryptosystems, lattice-based cryptography has attracted significant interest from the research community owing to its robust security guarantees and relatively low computational complexity~\cite{LWE,lattice_ency}. Lattice-based cryptographic algorithms rely on the mathematical concept of a lattice, which is an intricate structure formed by repeating patterns of points in a multi-dimensional space.


One of the fundamental building blocks of lattice-based cryptographic algorithms is polynomial operations, specifically, polynomial modular multiplication (PMM). PMM is a critical operation in ring learning with error (RLWE) theory, a key concept in lattice-based cryptographic algorithms. Moreover, PMM is the most time-consuming part of these algorithms. For example, recent studies show that PMM represents more than half of the computational workload for lattice-based homomorphic encryption (HE) on the cloud side~\cite{cheetah}, and more than 90\% on the edge side~\cite{CIM_HE_SAC}. Though algorithmic optimizations like Number-Theoretic Transform (NTT)~\cite{NTT} can decrease computation complexity, PMM latency is still high~\cite{cheetah,HE_F1,CIM_HE_SAC}. As such, accelerating PMM is essential to improve the efficiency and practicality of lattice-based cryptographic algorithms. 

Currently, there have been significant efforts to accelerate PMM, particularly through the use of NTT. NTT-based solutions, including those implemented on application-specific integrated circuits (ASICs)~\cite{asic_ntt,leia,sapphire}, field-programmable gate arrays (FPGAs)\cite{fpga_ntt}, and compute-in-memory (CIM) architectures\cite{rmntt,bpntt,CryptoPIM,MENTT}, have demonstrated promising results in accelerating PMM. CIM-based PMM accelerators have gained attention for their effectiveness in reducing data transfer overheads by moving computation inside the memory~\cite{CIM_HE_SAC,CryptoPIM}. Work in~\cite{CryptoPIM} builds a Resistive RAM (ReRAM) based NTT accelerator that supports bit-wise computation inside the memory. Alternatively, \cite{MENTT} presents an in-SRAM NTT accelerator with bit-serial arithmetic operations. Crossbar arrays (XBAs)~\cite{fundamental} is another popular CIM fabric that can support highly efficient vector-matrix multiplication (VMM) and is also actively being exploited for supporting high-throughput NTT-based PMM implementations~\cite{rmntt,iedm_ntt}. 

Existing research efforts to accelerate PMM using XBAs have primarily focused on using NTT-based approaches~\cite{rmntt, iedm_ntt}. Such solutions claim to achieve improvements of over 50$\times$ compared to other CIM NTT accelerators. However, supporting PMM on XBAs comes with its own set of unique challenges. These challenges differ notably from those associated with the application of XBAs for the well-studied case of convolutional neural networks (CNNs). On the one hand, the high bitwidth and the large polynomial degree required for cryptographic applications result in a huge number of shift-add operations, which incur high area and energy overhead. On the other hand, it remains an open question whether NTT-based solutions are the most suitable for XBA-based CIM architectures. Existing NTT-based PMM implementations on XBAs suffer from high area costs and limited scalability. These challenges restrict existing XBA-based solutions from achieving high performance for lattice-based cryptographic algorithms. Therefore, exploring alternative approaches to accelerate PMM to overcome these limitations is crucial. 




This paper proposes a novel XBA-based PMM accelerator, \name. Our solution distinguishes itself from existing XBA-based CIM methods by focusing on the non-NTT-based PMM. Our specific contributions are as follows:

\begin{itemize}
    \item We present observations revealing that NTT-based PMM may not be the most suitable choice for XBA-CIM. Our extensive studies show that the convolution 1D (Conv1D) solution holds potential advantages regarding area, latency, and noise over NTT when implementing PMM on XBAs.
    
    \item We propose a new XBA bit mapping technique for high-bitwidth, large polynomial degree data. The technique significantly reduces the overhead by removing most fine-grained shift-add operations. 
     \item We optimize data mapping at the processing engine (PE) level to support different problem scales with a fixed number of XBAs while maximizing throughput.
\end{itemize}

Our proposed \name offers significant improvements in throughput and area consumption, making it a competitive solution for accelerating PMM in lattice-based cryptographic algorithms. Specifically, \name achieves 200$\times$ latency improvement compared with a CPU implementation. It also leads to 3.9$\times$ throughput per area improvements compared with the state-of-the-art (SOTA) CIM accelerators for PMM. 

\section{Background}
\label{sec:background}
In this section, we discuss the role of PMM in cryptography, describe various PMM methods, review existing strategies for accelerating PMM, and review the concept of XBA.
\subsection{PMM in Cryptography}

The RLWE problem~\cite{RLWE}, foundational to lattice-based cryptography~\cite{lattice_theory}, and specifically to HE schemes~\cite{HEStandard}, leverages polynomials over a specific ring for its operations. HE, which enables arbitrary computations on encrypted data without prior decryption, ensures secure computation in untrusted environments while preserving data privacy. The primary computational bottleneck in HE arises from the need to perform polynomial arithmetic, particularly PMM~\cite{B/FV,BGV,CKKS}. Consequently, enhancing PMM's performance with respect to latency and energy consumption becomes critical in cryptography.

\subsection{PMM}\label{sec:pmm}

\edit{Polynomial modular multiplication (PMM) is a fundamental operation in various applications, including cryptography, error correction codes, and polynomial arithmetic. It involves multiplying two polynomials and reducing the result modulo a given polynomial, resulting in a polynomial of a lower degree. By performing PMM, it becomes possible to efficiently compute large polynomial expressions while maintaining the desired modulus properties.
}

\begin{figure}
    \centering
    \includegraphics[width=\linewidth]{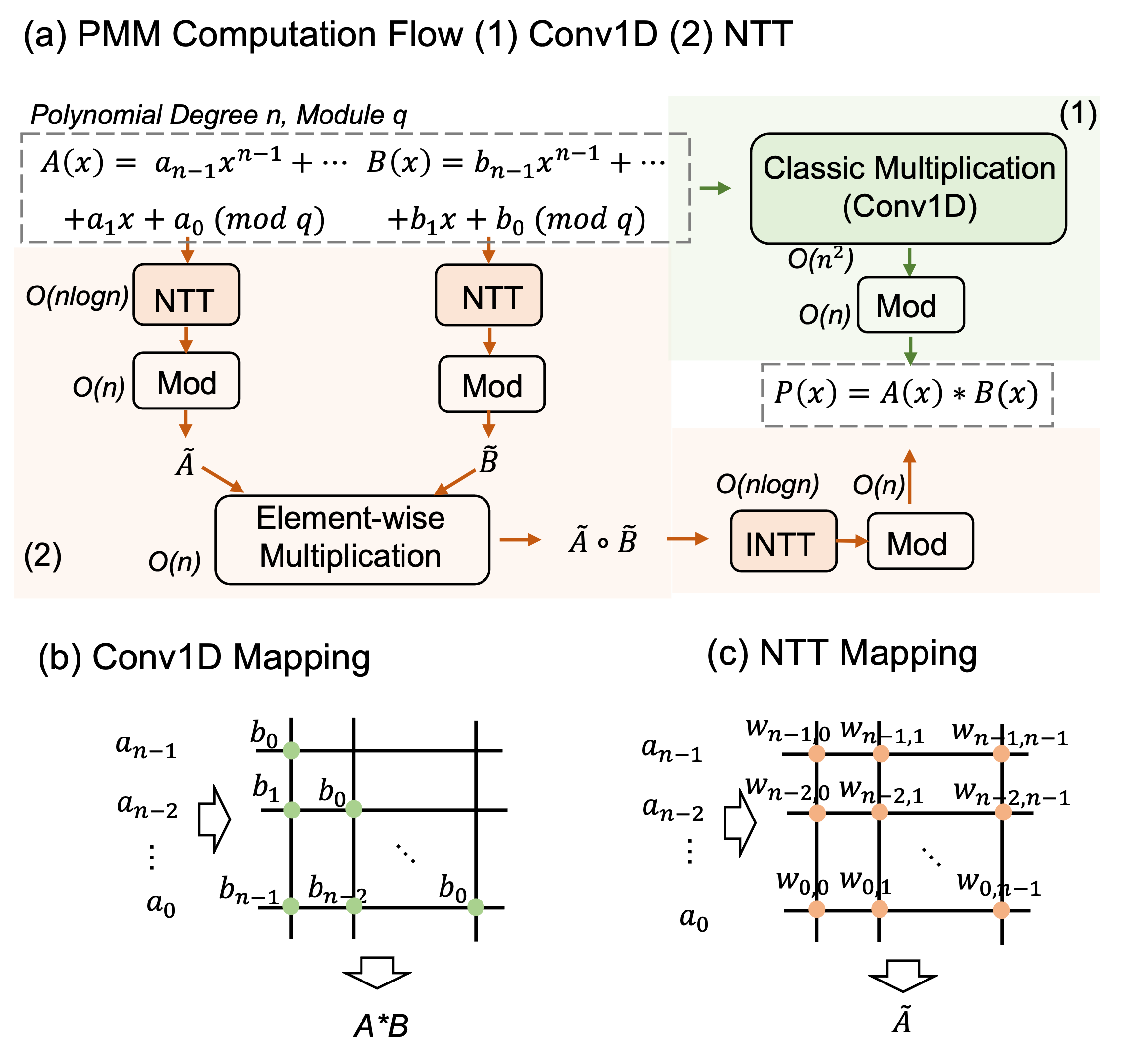}
    \caption{(a) PMM computation flow using two implementations: (1) Conv1D (2) NTT. (b) PMM operation mappings on XBA: Conv1D mapping; (c) NTT mapping.}
    \label{fig:pmm}
    \vspace*{-8mm}
\end{figure}

\edit{PMM can be accomplished using various methods, including the Conv1D approach and more optimized solutions like NTT as shown in Fig.~\ref{fig:pmm}(a). The Conv1D approach for PMM follows a straightforward procedure (Fig.~\ref{fig:pmm}(a)(1)). Two polynomials $A(x)$ and $B(x)$, with polynomial degree $n$ and modulo $q$,  are multiplied by summing the corresponding terms, akin to Conv1D computation with time complexity of $O(n^2)$. Then, the product undergoes modular reduction by dividing it with a modulus polynomial. The remainder is extracted polynomial long division to get the final result $P(x)$.}

\edit{NTT, alternatively, is proposed to reduce the computational complexity of PMM, particularly when the modulus polynomial satisfies specific properties, such as being irreducible and having a specific degree~\cite{NTT}. As depicted in Fig.~\ref{fig:pmm}(a)(2), the NTT approach involves transforming the polynomials into a different domain through NTT. During the NTT transformation, butterfly computations are performed by combining pairs of coefficients and multiplying them with twiddle factors, which are complex values associated with the modulus polynomial, resulting in the frequency-domain representation of the polynomial~\cite{NTT}. The process has a time complexity of $O(n\log{n})$. Then in this transformed domain, element-wise multiplication is performed, followed by the inverse NTT (INTT) to convert the result back to the original domain to obtain the final polynomial $P(x)$. Modular reduction is applied after each domain transformation.}

\edit{The computational complexity of PMM in hardware is primarily influenced by two key factors: the polynomial degree $n$, which represents the number of coefficients in a polynomial, and the bitwidth $k$ of modulo $q$, which signifies the size of these coefficients. In real-world applications, such as HE in privacy-preserving machine learning inference, these parameters can be quite substantial. For instance, the polynomial degree $n$ in these applications can range from 256 to 8192, while the bitwidth $k$  can vary from 16 bits to 64 bits \cite{GAZELLE,cheetah}. The magnitude of these degrees and bitwidths significantly intensifies the computational complexity of a single PMM, presenting a considerable challenge in the field.}

\subsection{Related Work}

In this section, we briefly review existing efforts to accelerate PMM. As existing work primarily employs NTT-based solutions, we focus our review accordingly, discussing both traditional ASIC and FPGA solutions, as well as CIM-based accelerators.

\subsubsection{ASIC and FPGA solutions}

 Nejatollahi, H., et al.~\cite{fpga_ntt} proposed an innovative FPGA solution by designing two high-throughput systolic array polynomial multipliers, one based on NTT and the other on convolution. Their sequential NTT-based multiplier yielded a 3$\times$ speedup over the SOTA FPGA implementation of the polynomial multiplier in the NewHope-Simple key exchange mechanism on an Artix7 FPGA \cite{newhope}. 

ASIC implementations of lattice-based cryptographic protocols have also been actively studied. LEIA~\cite{leia}, a high-performance lattice encryption instruction accelerator, and Sapphire~\cite{sapphire}, a configurable processor for low-power embedded devices, both demonstrate substantial performance improvements and energy efficiency compared to prior ASIC designs.

There are also a number of works that directly accelerate HE, inherently accelerating PMM~\cite{HE_F1,craterlake,BTS,cheetah,CIM_HE_SAC}. These works, which also typically use NTT, aim to create large-scale accelerators for privacy-preserving computations. Since this paper focuses on PMM, we will not compare it to these works. It suffices to say that an efficient PMM accelerator will directly help HE implementations.

\subsubsection{Compute-in-Memory solutions}
\edit{Previous research has introduced a variety of CIM kernels, including crossbars and general-purpose CIM. Ranjan et al.~\cite{ranjan2019x} have demonstrated that XBAs excel at performing VMM. Reis et al.~\cite{reis2018computing} have discussed the general-purpose CIM enabling Boolean logic and arithmetic operations to be executed directly within the memory.  Additionally, ongoing researches focus on exploring different underlying technologies for implementing these CIM kernels, including CMOS, ReRAM, and Ferroelectric FET (FeFET)~\cite{ni2019ferroelectric}. These technologies are actively studied due to their potential to provide higher density and lower latency/energy overhead in CIM architecture.}
Several research efforts have explored the use of CIM architectures for the acceleration of the NTT, including CryptoPIM~\cite{CryptoPIM}, MENTT~\cite{MENTT}, RMNTT~\cite{rmntt} and BPNTT~\cite{bpntt}. We compare \name against these established researches, so we concisely introduce these approaches in the following discussion.

CryptoPIM, MENTT, and BPNTT proposed efficient NTT accelerators based on general-purpose CIM kernels. CryptoPIM \cite{CryptoPIM} and MENTT \cite{MENTT}, built on ReRAM and SRAM respectively, both introduced unique mapping strategies to streamline the data flow between NTT stages, leading to significant reductions in latency, energy, and area overheads. BPNTT presented an in-SRAM architecture using bit-parallel modular multiplication, significantly improving throughput-per-watt. 

RMNTT~\cite{rmntt} proposed an NTT accelerator using ReRAM-based XBAs. RMNTT stores the modified twiddle factor matrix in the XBAs and employs a modified Montgomery reduction algorithm to perform modular reduction on the VMM results. The evaluation results in~\cite{rmntt} show that RMNTT outperforms other NTT accelerators in terms of throughput but incurs a large area overhead.

\subsection{Crossbars} \label{sec:Crossbar}
Given the competitiveness of XBA-based NTT accelerators, we consider leveraging XBAs to accelerate PMM. We briefly review the XBA basics below. 

XBA~\cite{SWIPE} is one representative CIM kernel in which every input signal is connected to every output signal through their cross-points consisting of memory elements and selectors. XBAs can efficiently implement VMM and have been widely studied for CNNs. In particular, XBA implemented with nonvolatile memory (NVM) devices such as ReRAM~\cite{wu2018methodology} have gained popularity due to their high storage density, nonvolatility, and low energy consumption. \edit{However, XBAs face challenges stemming from the underlying memory devices and circuits. In-situ memory device nonidealities, e.g., non-linearity, thermal noise, and variations, impact computed accuracy.}

\begin{figure}
    \centering
    \includegraphics[width=\linewidth]{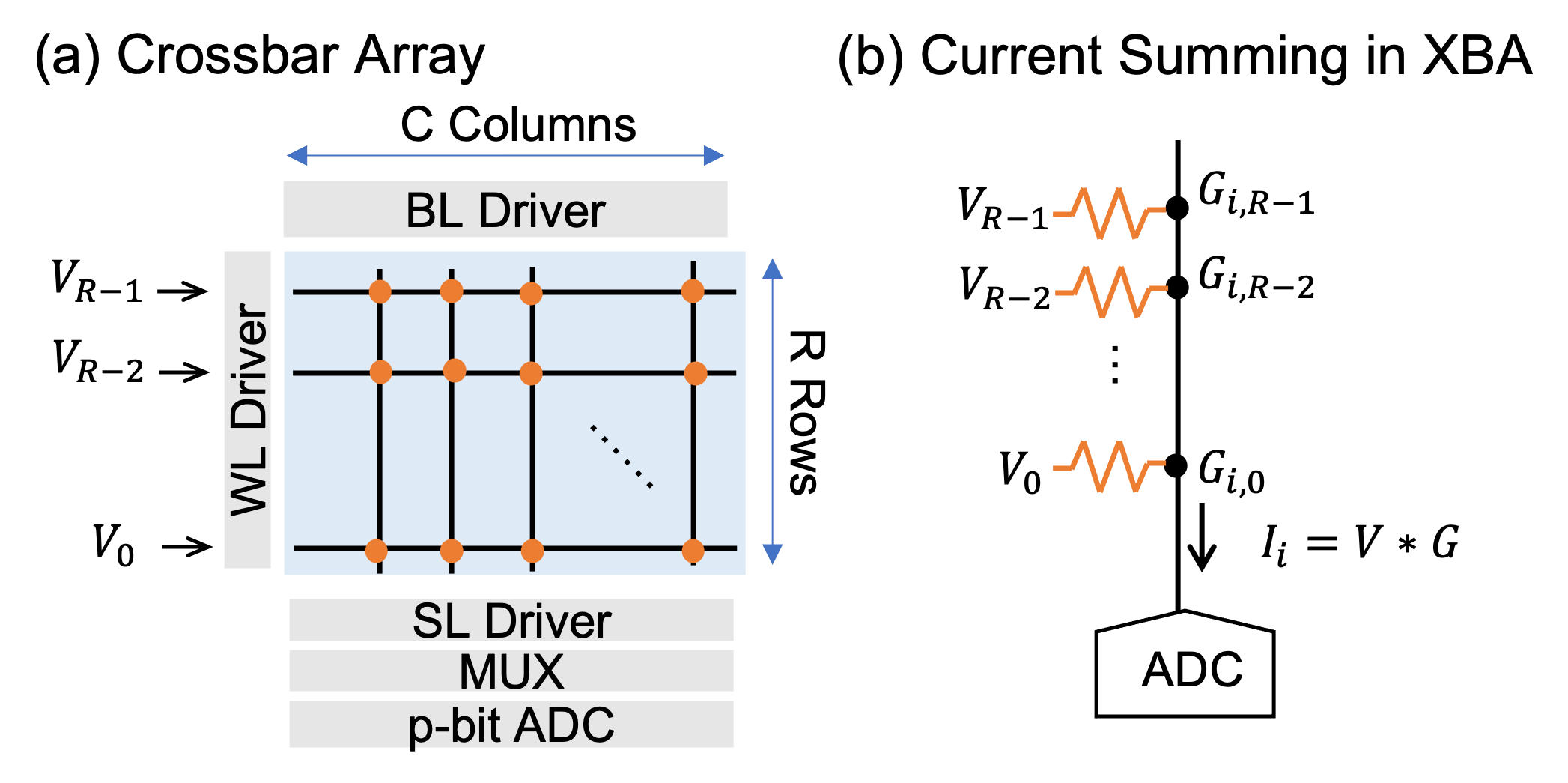}
    \caption{(a) XBA structure: a C columns x R rows array and the corresponding WL/BL driver. A p-bit ADC is used for converting analog signals to digital signals. (b) Illustration of the current summing scheme in XBA computation. }
    \label{fig:crossbar}
\end{figure}

Fig.~\ref{fig:crossbar}(a) illustrates a general XBA structure. For each column, we adopt the current summing model as shown in Fig.~\ref{fig:crossbar}(b). In this work, both input voltage ($V_{j}$) and memory cell states ($G_{i,j}$) assume binary values, i.e., $I_{i} = \sum_{0}^{R-1}G_{ij}V_{j}$, where $V_{j}$ and $G_{ij}$ are either 0 or 1. Binary XBAs exhibit greater robustness to device and circuit nonidealities, and offer improved scalability.

\section{NTT vs. Conv1D} \label{sec:discussion}

The choice of PMM algorithm is critical to achieving high performance in terms of speed, noise, and area in the context of the CIM computing paradigm as discussed in Sec.~\ref{sec:pmm}. Two commonly used methods for performing PMM are Conv1D and  NTT. Recent efforts utilizing XBAs for PMM have primarily focused on accelerating NTT-based methods~\cite{rmntt}~\cite{iedm_ntt}. However, there is no systematic comparative study of which method, Conv1D or NTT, is a better fit for leveraging XBAs to accelerate PMM. We fill this gap with an in-depth investigation below. Our study reveals three key insights which favor the Conv1D over the NTT-based approach. First,  data mapping complexity is higher when using NTT. Second, the Conv1D method potentially offers a better performance trade-off in terms of area and throughput, providing more opportunities for design scalability. Third, the noise growth is generally higher in the NTT approach than Conv1D, which can negatively impact the performance and accuracy of the system. Below, we elaborate on these insights.

\subsection{The Impact of Data Mapping to XBAs}


To use XBAs for PMM, both NTT-based and Conv1D-based PMM approaches require converting their respective operands into matrices and performing VMM on XBAs\cite{rmntt}. In the NTT-based approach, the twiddle factor of NTT must be converted into a matrix. In the Conv1D approach, one of the polynomials is transformed into a matrix, while the other remains a vector, facilitating the execution of VMM. Data mapping to XBAs in NTT and Conv1D can be better visualized in Fig.\ref{fig:pmm}(b) and (c). The figures show that the same number of memory cells are needed for both methods; thus, NTT does not provide benefits over Conv1D in terms of the XBA area. Also, due to the butterfly computation involved in NTT, converting the twiddle factors into a matrix is significantly more complex than converting a polynomial into a matrix for Conv1D~\cite{rmntt}. 

The end-to-end computational complexity of NTT-based PMM on XBAs is actively higher than directly mapping Conv1D into XBAs. As depicted in Fig.~\ref{fig:pmm}(a), NTT-based PMM involves three main steps: NTT computation ($O(n\log n)$ complexity), element-wise multiplication ($O(n)$ complexity), and INTT computation ($O(n\log n)$ complexity). In contrast, Conv1D-based PMM has a complexity of $O(n^{2})$. However, when utilizing XBA acceleration, the complexity of Conv1D-based PMM can be reduced from $O(n^{2})$ to $O(1)$. By employing similar data mappings, NTT and Conv1D exhibit the same time complexity on XBA. Therefore, Conv1D-based PMM on XBA demonstrates a lower end-to-end complexity compared to NTT-based PMM, as it requires fewer operations—Conv1D only necessitates $O(1)$ operations, while NTT involves $O(1)$ + $O(n)$ + $O(1)$ operations.

\subsection{Performance Analysis}\label{sec:Performance_analysis}

NTT-based PMM requires that the twiddle factors be stored for NTT and INTT in the XBAs~(See Fig~\ref{fig:pmm}(c)). The stored twiddle factors approach  necessitates either frequent updates to the twiddle factors stored in the XBAs or the use of additional XBAs to store all twiddle factors needed for NTT. As a result, this leads to either higher latency and energy consumption or increased area. Alternatively, Conv1D-based PMM has numerous identical values that, when stored in XBAs, can be reused repeatedly. This provides the opportunity to devise intelligent data reuse schemes (see Sec. \ref{sec:poly_mapping}), ultimately leading to more efficient and optimized solutions in terms of area and energy consumption. Therefore, Conv1D-based PMM can be a more promising method for accelerating PMM with XBAs.

\subsection{Noise}

As discussed in Sec~\ref{sec:Crossbar}, XBAs are susceptible to accuracy degradation stemming from the intrinsic nonidealities of the memory cells, and the limitation of ADC precision. As a result, using XBAs inevitably introduces a certain amount of noise (i.e., error) in VMM results. When implemented on XBAs, Conv1D-based PMM incurs less noise than NTT-based PMM. The primary reason is that in Conv1D-based PMM, the entire computation can be completed in one step in XBAs, which helps control the magnitude of the noise. However, in NTT-based PMM, the NTT, element-wise multiplication, and INTT must be performed, which increases the noise introduced by XBAs multiplicatively (See Fig~\ref{fig:pmm}(a)). In applications such as HE, higher noise levels are not tolerable, making NTT-based XBA PMM unsuitable for such applications.

Based on the observations in this section, we believe that Conv1D-based PMM is a better approach for accelerating PMM with XBAs. We thus focus on the design and optimization of the XBA fabric to accelerate Conv1D-based PMM.

\section{\name}\label{sec:design}
\eat{
\begin{figure*}[thb]
    \centering
    \includegraphics[width=0.95\linewidth]{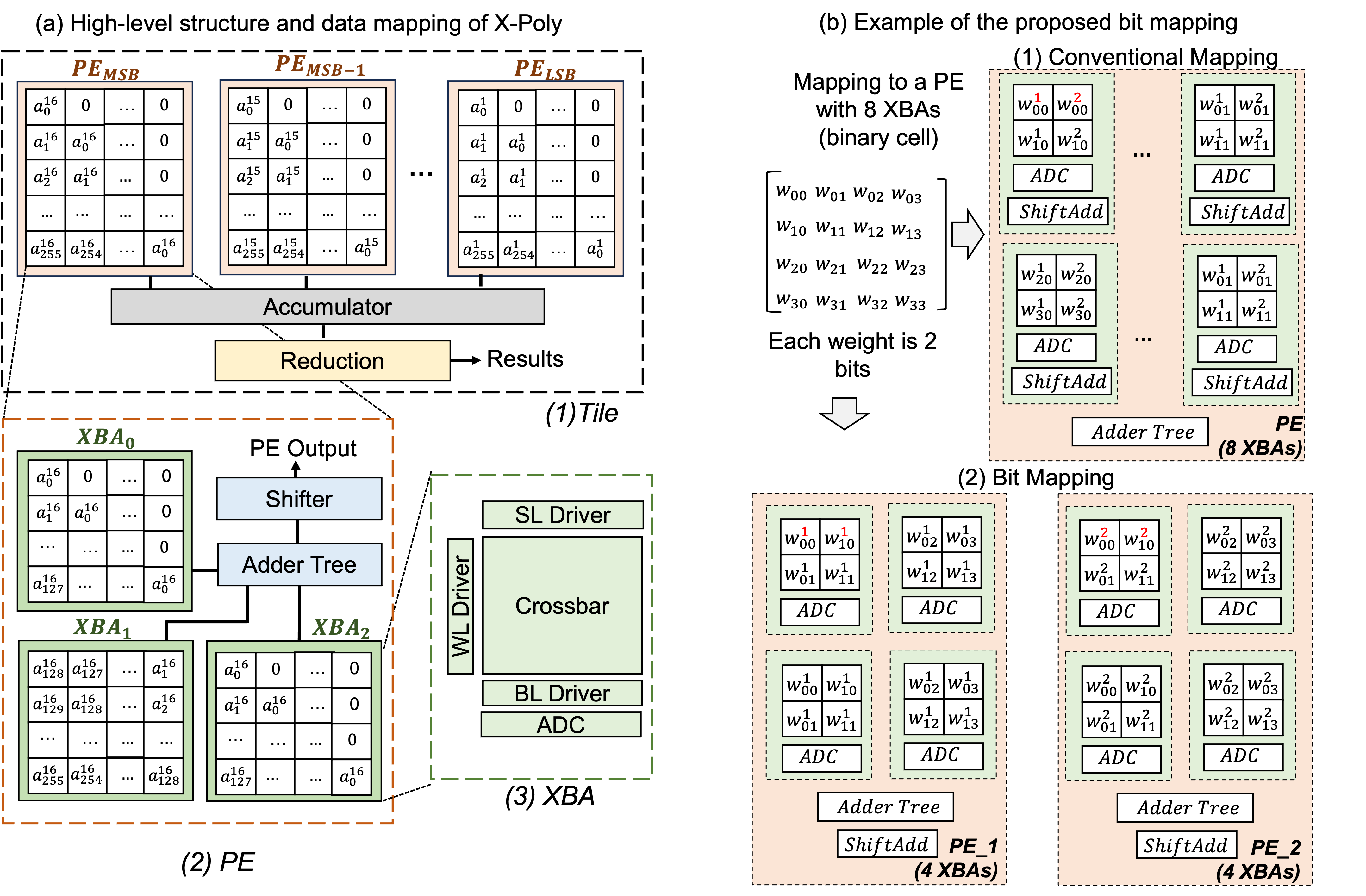}
    \caption{(a) Hierarchial structure of the Proposed \name design: (1) Tile level design and data mapping, (2) PE level design and data mapping, and (3) XBA structure. (b) Example of the proposed bit mapping technique: mapping 2bit 4x4 weights to 2x2 XBA with binary cells: (a) Conventional mapping. (b) Bit mapping.}
    \label{fig:overall}
\end{figure*}}

\begin{figure}[t]
    \centering
    \includegraphics[width=\linewidth]{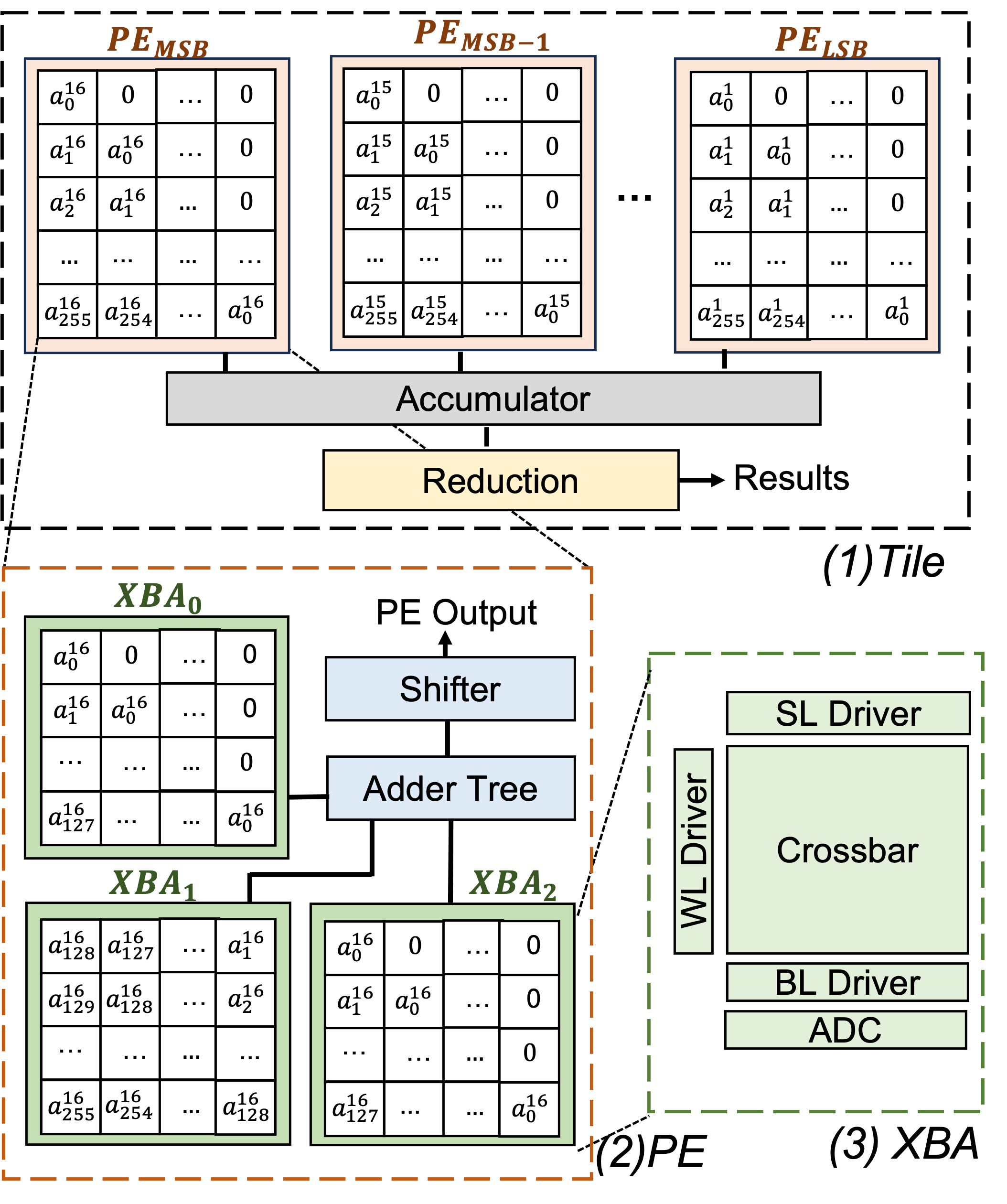}
    \caption{Hierarchical structure of the proposed \name design: (1) Tile level design and data mapping, (2) PE level design and data mapping, and (3) XBA structure.}
    \label{fig:overall}
    \vspace*{-5mm}
\end{figure}

\edit{Design and optimization of Conv1D-based PMM on XBAs  for long polynomials must solve several key problems. These include mapping data to XBAs to efficiently  use the resources, enhancing memory utilization at the Processing Element (PE) level, and effectively implementing modular reduction strategies. We present \name for accelerating the Conv1D-based PMM and provide tailored solutions to address the aforementioned challenges.}



\subsection{Overview}\label{sec:overview}

The high-bitwidth long polynomials employed in cryptographic algorithms like HE propose challenges for the design of XBA-based architecture. One specific issue relates to the limited size of the XBA. For instance, an array with 128 rows and 128 columns falls short in accommodating high-bitwidth polynomials with a degree exceeding 256.

To address the challenge, \name utilizes a hierarchical approach to address computational complexity. Fig.~\ref{fig:overall} illustrates the overall structure and data mapping of \name, consisting of the tile, PEs, and XBAs. The tile (Fig.~\ref{fig:overall}(1)) contains multiple PEs, an accumulator, and a specifically designed reduction unit for modular reduction. Each PE holds one-bit weights and shares the same input. Thus, $k$ PEs can store $k$-bit polynomials from the most significant bit (MSB) to the least significant bit (LSB), working in parallel.

The PE (Fig.~\ref{fig:overall}(2)) is composed of multiple XBAs working on different parts of the polynomials simultaneously, as well as an adder tree and a shifter. The XBAs (Fig.~\ref{fig:overall}(3)) are used for coefficient multiplication, while the adder tree and shifter within each PE accumulate partial results from each XBA and perform shift-add operations.

\subsection{Bit Mapping}\label{sec:mapping}

The high bitwidth and large polynomial degree required for cryptographic applications need a large number of shift-add operations, which may not be efficiently supported in a CIM architecture. Due to the limited precision of a memory cell in an XBA, we need to map the bits of weight into multiple memory cells. Fig.~\ref{fig:mapping}(a) illustrates the conventional approach for mapping the high bitwidth weight to multiple XBAs. All bits of weight are stored in multiple columns of the XBA. When input arrives at the XBA, each column conducts a multiplication operation. Immediately following this, shift-adders carry out the shift-add operations after the XBA computation. This XBA-level shift-add operation requires lots of shift-adders and is expensive in terms of both time and energy. 

As such, in this work, we propose a new bit mapping (BM) technique that groups the same bit of all weights together, as shown in Fig.~\ref{fig:mapping}(b). For example, in the case of 4x4 2-bit weights distributed among 2 PEs (4 XBAs per PE), each PE process one bit of each weight. After all PEs process one input bit, the shift operation is performed at the PE level, thereby avoiding a costly array-level shift-add operation. Comparing the conventional mapping (Fig.~\ref{fig:mapping}(a)) and the bit mapping (Fig.~\ref{fig:mapping}(b)) in the example, the number of shift-adders is reduced from 8 to 2.

\begin{figure}[thb]
    \centering
    \includegraphics[width=0.95\linewidth]{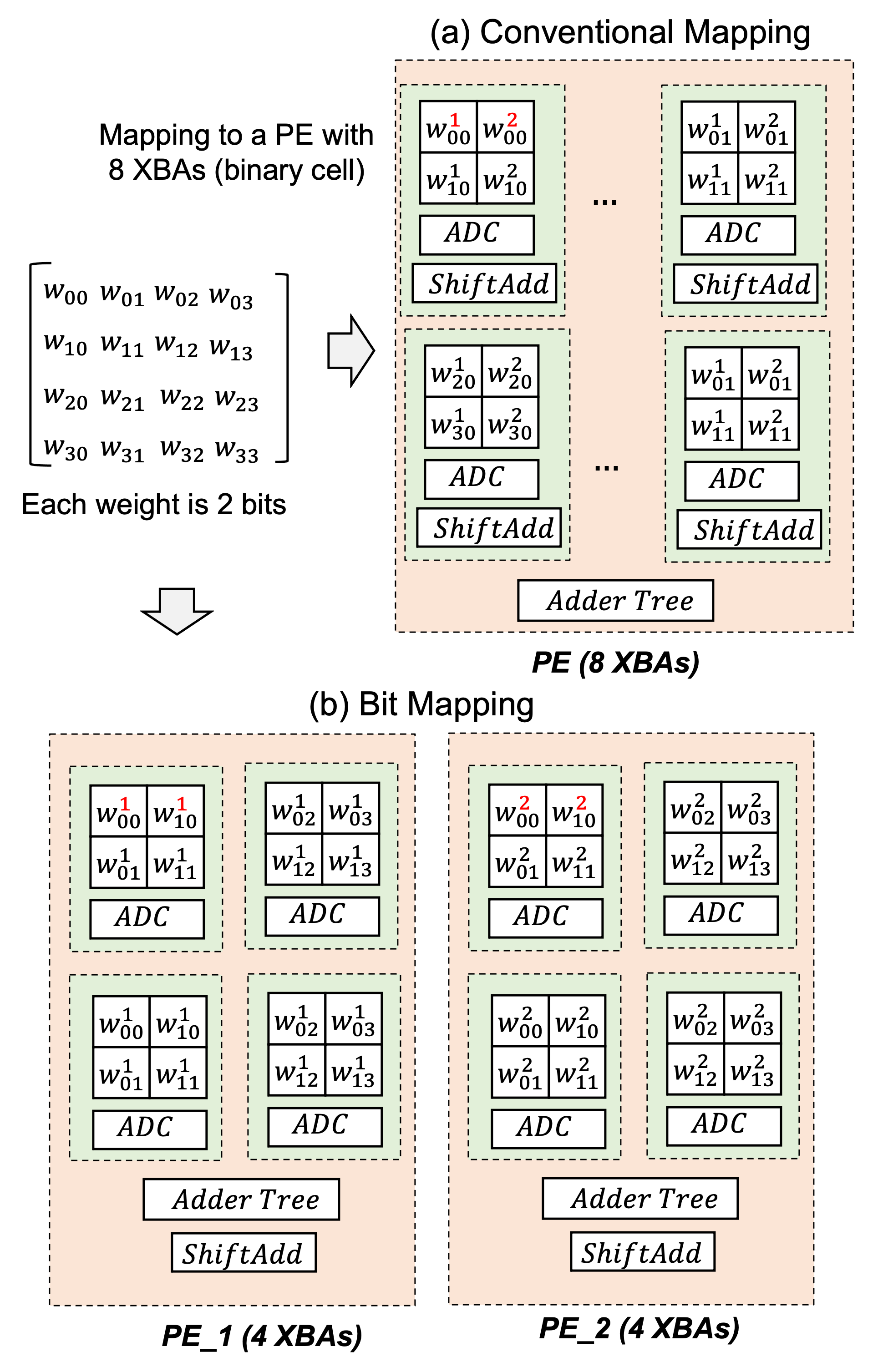}
    \caption{Example of the proposed bit mapping technique: mapping 2bit 4x4 weights to 2x2 XBA with binary cells: (a) Conventional mapping. (b) Bit mapping.}
    \label{fig:mapping}
    \vspace*{-5mm}
\end{figure}

As will be seen, this bit mapping strategy can significantly improve both the area and speed for processing high-bitwidth polynomial-based workloads in XBAs. In addition to its benefits for shift operations, the BM technique also simplifies the design of the PE. Since each PE handles a bit of each polynomial, the data patterns are captured at the polynomial coefficient level. We can simultaneously perform mapping optimization for all PEs. Thus, this technique can be easily extended to accommodate polynomials with different degrees or bitwidths, making it a flexible solution for performing polynomial operations in XBAs.

\eat{
\begin{figure}[thb]
    \centering
    \includegraphics[width=\linewidth]{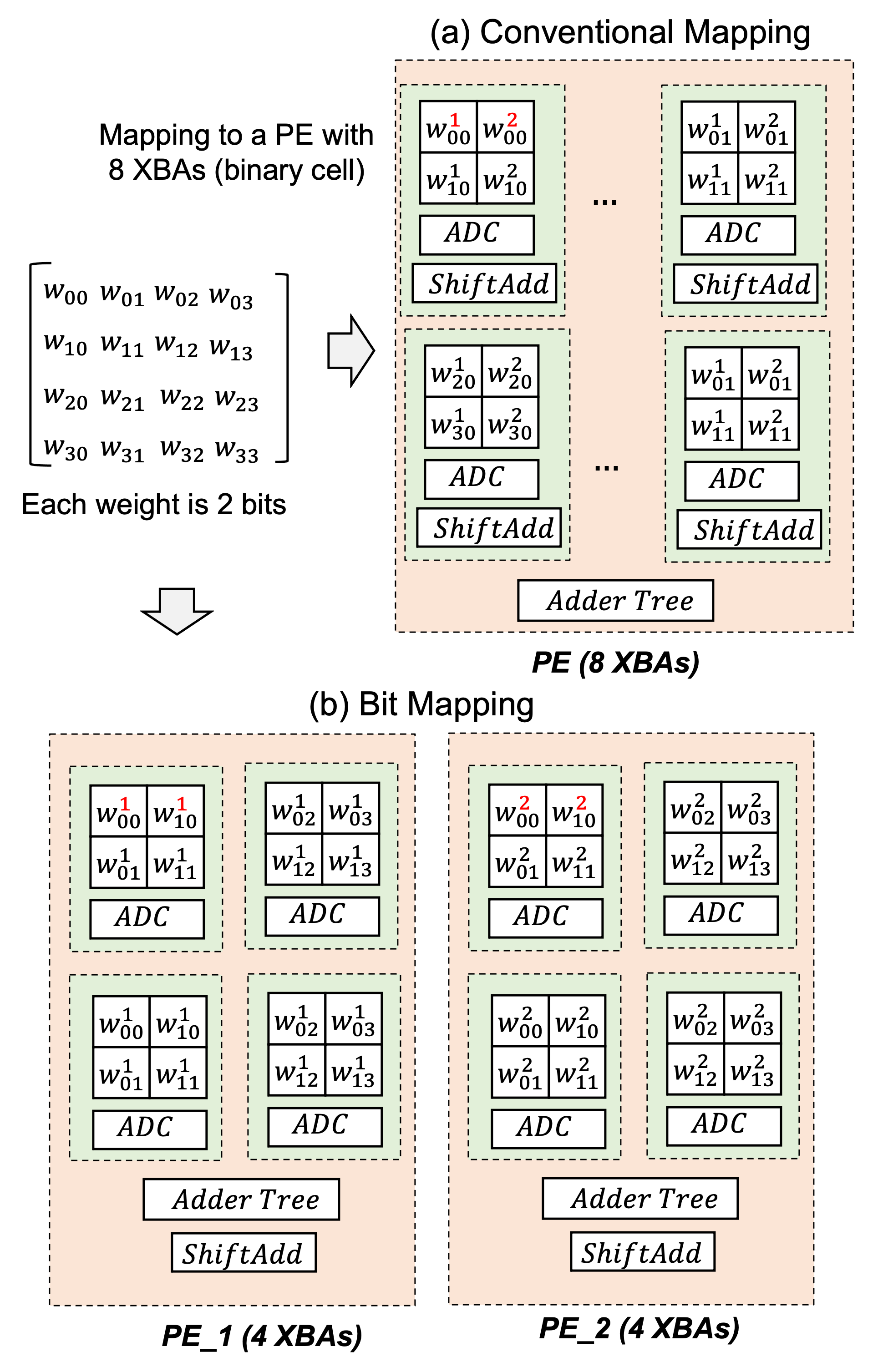}
    \caption{Mapping 2bit weights to 2x2 XBA with binary cells: (a) Conventional mapping. (b) Bit mapping.}
    \label{fig:mapping}
    \vspace*{-5mm}
\end{figure}
}

\subsection{Polynomial Mapping}\label{sec:poly_mapping}

In our PMM approach (utilizing VMM in XBAs), we first map polynomials into matrices to facilitate computation. Each polynomial is converted into a matrix by horizontally shifting the coefficients of the polynomial across each row, with any remaining gaps filled with zeros. This procedure results in a matrix structure that supports the critical shift-add operations intrinsic to PMM. 

Mapping the matrices into XBAs in our PMM approach using VMM is straightforward. However, in an effort to further optimize this mapping scheme, we noted that for any given polynomial degree $n$ and XBA row length $x$, there is a consistent pattern of repeated XBAs. Specifically, in every instance, we require $n/x$ identical XBAs to represent the polynomial matrix.

This aspect of our design stands in contrast with NTT-based XBA designs~\cite{rmntt}, which often find themselves confined to specific polynomial parameter settings. As such, \name offers a significant increase in flexibility. For example, consider two application scenarios for privacy-preserving machine learning (PPML) inference as shown in~\cite{GAZELLE}. On server-side inference, where performance is prioritized, and energy or area constraints are less critical, \name can leverage a larger count of XBAs for high-throughput PMM in HE of PPML. For edge-device inference, where area and energy efficiency are paramount, \name can efficiently handle a variety of large polynomial degrees and bitwidths with a smaller number of XBAs. A detailed study of the scalability of our design, referred to as \name, is provided in Sec. \ref{sec:memory_utilized}.

\subsection{Modular Reduction}\label{sec:modular}
Modular reduction is a crucial step in PMM, which ensures that the resulting polynomial remains within a specified degree and coefficient bounds. The essential steps in the reduction process include selecting an appropriate modulus for the ring and performing the modulo operation on the degree and coefficients of the resulting polynomial.

In \name, we utilize a variant of the Barrett reduction~\cite{Barrett} technique for efficient modular reduction. This method is known for its effectiveness in cryptographic applications and modular arithmetic, as it can compute the remainder of a division operation without performing the division itself. Notably, to minimize computation overhead in reduction, we strategically pre-compute specific parameters. This strategy transforms the complex, time-consuming multiplication and division operations into  shift operations, effectively reducing computation time and optimizing the overall reduction process.

\subsection{Computation Flow}\label{sec:flow}

Assuming polynomial $A$ is mapped onto XBAs in \name, PMM can be accomplished as follows. (1) \textbf{Input processing}: We begin by bit-slicing each element in the new polynomial B, separating it into its individual bits. (2) \textbf{PE computation:} Within each PE, different arrays handle distinct sections of the polynomial and perform multiplications with corresponding sections of the input. The results are then summed and shifted at the PE level. This bit-by-bit input process continues until all input bits have been addressed. (3) \textbf{Tile accumulation:} Afterward, the results from all PEs are accumulated at the tile level, and the partial results obtained from each PE are combined. (4) \textbf{Tile reduction:} Finally, a tile-level reduction operation is applied for efficient modular reduction. 

We also prioritize maximizing throughput in our design by incorporating a three-stage pipeline into the \name workflow to enhance the PMM process. This pipeline, which encompasses the PE computation, tile accumulation, and tile reduction stages, enables efficient synchronization and overlapping operations. 

\section{Evaluation}
\label{sec:evaluation}

In this section, we present the evaluation of \name. We begin by discussing our implementation setup and evaluation tools and follow with a comparison with both the CPU-based solutions as well as other hardware accelerators. We will quantitatively assess the performance benefits from \name. We then evaluate our bit mapping technique, with a focus on energy and area savings. Then we study the throughput per area performance of \name, demonstrating its superior performance over other SOTA CIM accelerators. Finally, we assess the scalability of \name and highlight its versatility in handling diverse polynomial degrees and bitwidths.

\subsection{Implementation Setup}
To verify the functionality of \name and evaluate performance characteristics such as latency, energy, and area, we have assembled a comprehensive evaluation framework. 

This framework considers the simulation of hardware components, including the modular reduction unit, shift-adders, accumulators and XBA arrays. We implemented the reduction unit, shift-adder, and accumulator using RTL, coded in Verilog and evaluated the energy consumption and area of these components using the RTL synthesis tool Cadence Encounter, paired with the 45nm CMOS predictive technology model (PTM) \cite{cao2011predictive}. We used Neurosim ~\cite{lu2021neurosim} to estimate the latency, energy, and area of the ReRAM-based XBAs, as well as successive-approximation-register (SAR) ADCs assuming the same 45nm technology node. The size of each XBA is 128 rows $\times$ 128 columns and one ADC is shared by 8 columns.

We then incorporated the aftermentioned simulation-based results into our Python-based cycle-accurate simulator. This simulator tracks the pipeline stages for a given PMM operation and computes the cycle count and total energy consumption by emulating the operations of each hardware component on a cycle-by-cycle basis. This evaluation framework allows us to generate a holistic and precise assessment of the overall performance of a PMM in \name.

\begin{table*}[b]
\centering
\caption{Comparison between \name and other SOTA solutions on a 256-point polynomial. Technology size: 45nm}
\label{tab:results}
\begin{tabular}{c|c|c|c|c|c|c|c|c|c c}
\hline 
\hline
& \multicolumn{2}{c|}{PMM Solutions} & \multicolumn{4}{c|}{NTT Solutions (CIM)} & \multicolumn{3}{c}{NTT Solutions (Non-CIM)} \\\hline

Design   & \textbf{\name} & \textbf{CPU} & \textbf{RMNTT} & \textbf{BPNTT} & \textbf{MENTT}& \textbf{CryptoPIM} & \textbf{FPGA}& \textbf{LEIA} & \textbf{Sapphire}\\ 
\hline
Device   &   ReRAM & CMOS & ReRAM & SRAM & SRAM & ReRAM & CMOS & CMOS  & CMOS\\ 
Frequency (MHz) & 400 & 2.5k & 400 & 3.8K & 218 & 909 & 164 & 267 & 64\\
Bit width & 16 & 16 & 14 & 16 & 14 & 16 & 16 & 14 & 14\\
\hline
Area ($mm^2$) & \textbf{0.27}  & - & $0.76^{*}$ & 0.063 & 0.173 & 0.152 & - & 1.77 & 0.354 \\
Latency ($us$) & \textbf{0.32} & 56 & 0.44 & 61.9 & 15.9 & 68.7 & 24.3 & 0.6 & 20.1\\
Energy ($nJ$) & 308.07 & - & 429.91 & 69.4 & 47.8 & 2.6k & 3.1k & 44.1 & 236.3\\
Throughput\$ (KOP$/s$) & $3.1k$ & - & $2.2k$  & $258.6$ & $62.8$ & $553.3$ & $41.2$ & $1.7k$ & $49.7$   \\
Throughput/Area\$ (KOP$/s/mm^2$) & \textbf{11.4k} & - & $2.9k$  & $4.1k$ & $364$ & $3.6k$ & - & $940.6$ & $140.1$  \\
\hline
\hline
  \multicolumn{11}{@{}p{\linewidth}@{}}{\small * We estimate the area for RMNTT based on the information reported in the paper. We utilize the same XBA area and peripheral components as \name for the sake of comparison.} \\
  \multicolumn{11}{@{}p{\linewidth}@{}}{\small \$ We evaluate the throughput based on the type of operations performed in the corresponding accelerators. We report the throughput of PMM for both X-Poly and CPU as well as the throughput of NTT for other accelerators as reported in the literature.}
\end{tabular}

\end{table*} 
 
\subsection{Comparison with SOTA Solutions}
\subsubsection{Comparison with CPU} We first compared our \name implementation with a CPU implementation that performs PMM with a SOTA C++ library (Number Theory Library version 11.5.1 \cite{ntl}). An Intel(R) Xeon(R) CPU E5-2680 v3 operating at 2.50GHz was used for the CPU implementation. The results are shown in Table~\ref{tab:results} (col 3). The latency of the \name design is 200$\times$ better than the CPU implementation. Performance enhancement is primarily due to the parallel compute capability and fast multiplication inherent in the XBAs in our CIM-based architecture, allowing for a much more efficient PMM execution.

\subsubsection{Comparison with other accelerators} Next, we compared \name with other SOTA accelerators. As current accelerators for PMM only use NTT, we compare our approach to SOTA accelerators that support NTT given a polynomial degree of 256. That said, NTT solutions require additional multiplications and the INTT to obtain final PMM results. Compared to \name, this may increase overall latency and energy consumption by 2$\times$. Moreover, with \name, we can generate PMM results in a single step without the need for additional multiplication or INTT.

\textbf{XBA solutions:} We first compared our implementation with other CIM solutions, specifically with ReRAM implementations. \edit{We scaled the latency and energy of~\cite{rmntt} to 45nm for a fair comparison to \name, following the methodology outlined in~\cite{bpntt}. Given that the study in~\cite{rmntt} did not provide area results, we carried out an estimation using the mapping methodology introduced in their publication. We assume the same area for XBAs and peripheral components as \name.}  Although our design exhibited similar latency to RMNTT, our improved mapping technique results in a significantly reduced area. That is mainly because we reduce the footprint of shift-adders to just 20\% of the original area by using the proposed BM technique, thereby leading to a 3.9$\times$ improvement in the throughput-per-area ratio. 

\textbf{Compute in SRAM solutions:} We also compared our implementation with in-SRAM solutions. The \name approach also improves throughput and throughput-per-area. Again, this can be attributed to both the parallel computing capability and the fast multiplication feature of our XBA-type mapping technique, which enables us to perform multiple computations simultaneously. More specifically, throughput is improved by 11$\times$, and throughput-per-area is improved by up to 3$\times$.

\textbf{Non-CIM solutions:} Finally, we compared \name with non-CIM solutions. The \name design is advantageous as it can store entire polynomial coefficients inside the XBAs. This feature eliminated the need for frequent access to on-chip memory for coefficients in long polynomials, which reduces data movement between the computing unit and the on-chip memory. This results in a reduction in both latency and energy consumption. Overall, \name outperformed ASIC and FPGA solutions in terms of throughput and energy efficiency. Compared to SOTA FPGA implementations, \name can achieve a remarkable 75$\times$ throughput improvement. When compared against SOTA ASIC implementations, \name can achieve a 2$\times$ throughput improvement.

\subsection{Bit Mapping Study}
\begin{figure}
    \centering
    \includegraphics[width=\linewidth]{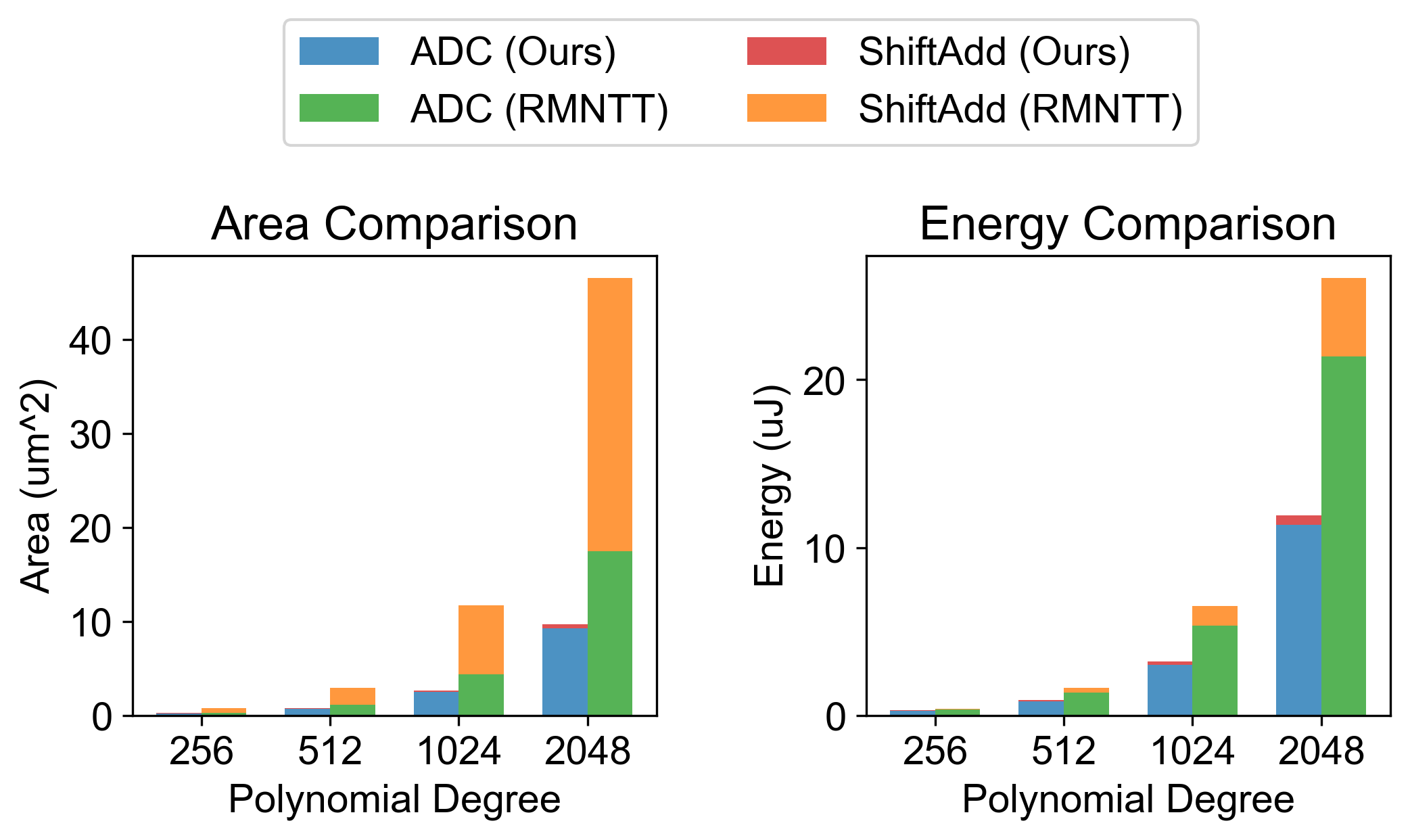}
    \caption{Comparison of area and energy breakdown for ADCs and shift-adders in \name and RMNTT~\cite{rmntt}.}
    \label{fig:MappingBenefit}
\end{figure}

We now evaluate the energy and area benefits of the proposed BM technique, discussed in Sec.\ref{sec:design}. To evaluate performance, we consider two scenarios: (1) conventional mapping as the implementation of RMNTT, the SOTA XBA-based NTT accelerator, and (2) our proposed BM technique. Fig.\ref{fig:MappingBenefit} illustrates the shift-adder area/energy and ADC area/energy given various polynomial degrees for each mapping. Results suggest that due to the large polynomial degrees and high bitwidths associated with the PMM, the peripherals (such as the shift-adders) in the design with conventional mapping consume a significant proportion of the energy and area. Moreover, this escalates with polynomial degrees. However, our proposed BM technique decreases the area for shift-add operations by 80\%, leading to an additional 3$\times$ reduction in overall area. Moreover, compared to conventional mapping, our design has lower latency and energy consumption. 

\begin{figure}
    \centering
    \includegraphics[width=\linewidth]{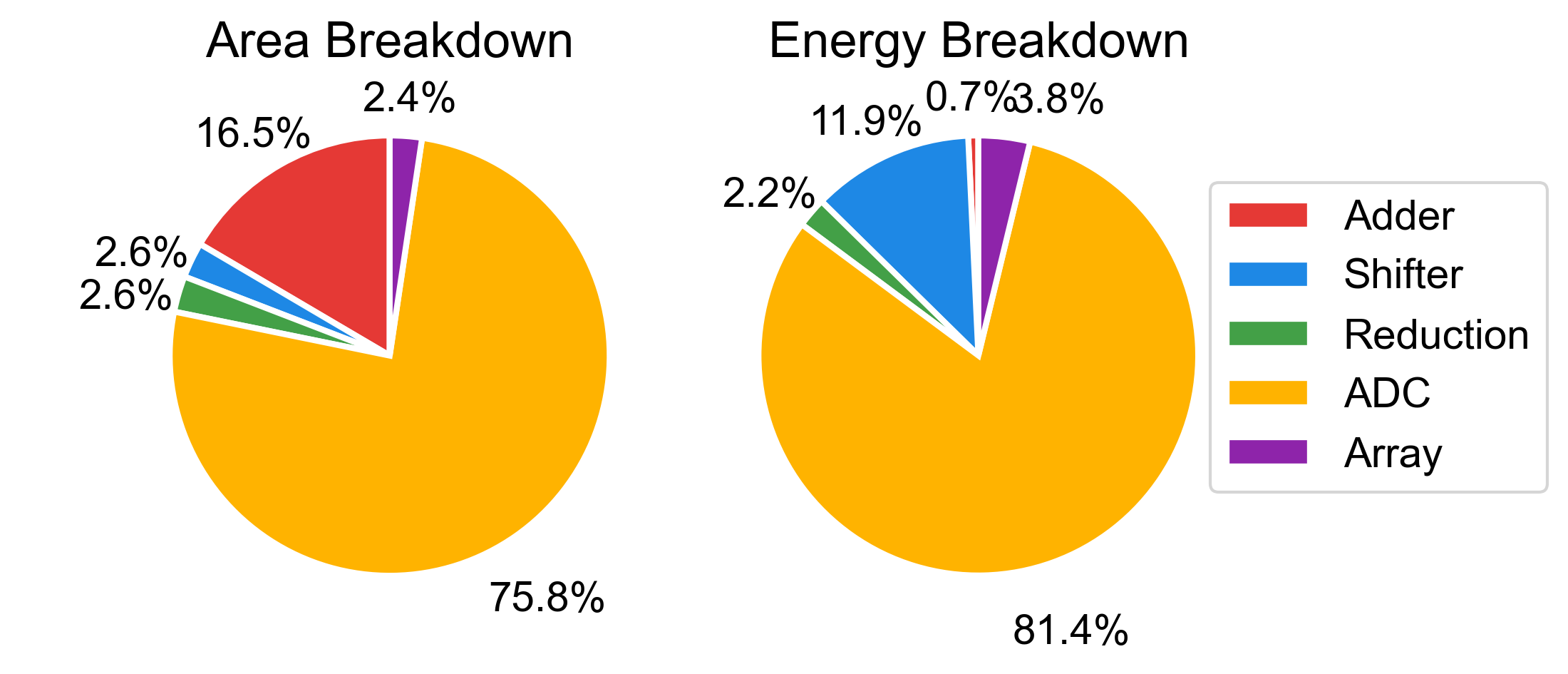}
    \caption{Area and energy breakdown for components of \name with polynomial degree 256 and bitwidth 16.}
    \label{fig:breakdown}
\end{figure}

Fig.~\ref{fig:breakdown} illustrates the area and energy breakdown of our proposed design. This analysis further reveals that the majority of the energy consumption and area is spent on ADC operations, with the proposed mapping technique reducing the energy and area consumption for other peripherals significantly. 

\subsection{Throughput per Area Study}
Table~\ref{tab:results} shows that the XBA-based solutions (\name and \cite{rmntt}) achieve higher throughput but require a larger area than the in-SRAM solution~\cite{bpntt}. This is due to the inherent design of XBA-based solutions: they require a more expansive area to accommodate an increase in both polynomial degree and bitwidth~\cite{rmntt}. In-SRAM solutions can support larger parameter sizes within a similar area. However, this is accompanied by a substantial reduction in throughput. However, \name reduces XBA area  while maintaining its high throughput. 

To further understand the trade-off between throughput and area, we conducted an analysis of the throughput per area performance and compare the results with other SOTA CIM solutions. We consider a range of polynomial degrees and bitwidths, to generate a comprehensive perspective regarding the strengths of our design.

\begin{figure}
    \centering
    \includegraphics[width=\linewidth]{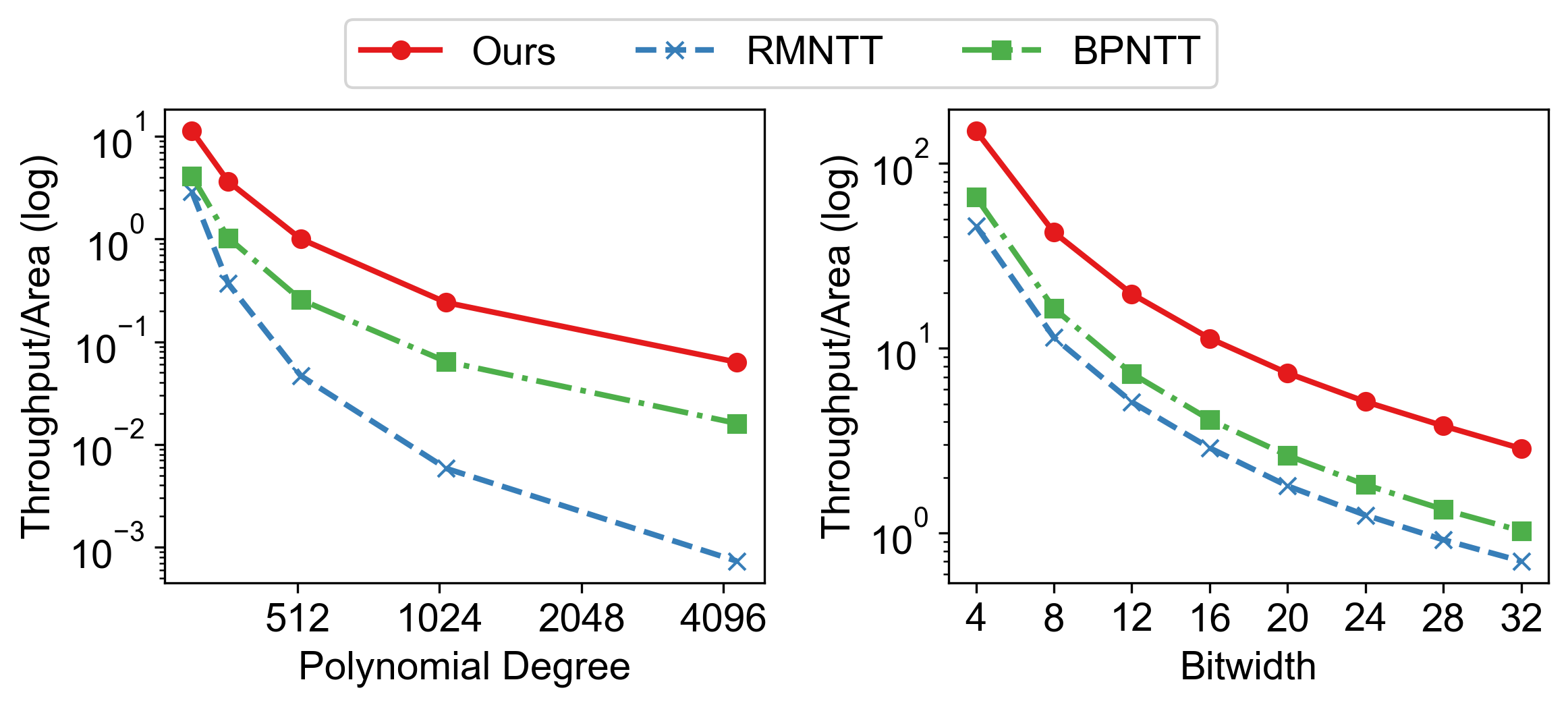}
    \caption{Throughput per Area (KOP/s/$mm2$) comparison with the SOTA CIM solutions (RMNTT and BPNTT) under different polynomial degrees and bitwidths. Y-axis is using log-scale for better illustration.}
    \label{fig:throughputarea}
\end{figure}

Fig.~\ref{fig:throughputarea} illustrates the throughput per area performance of our design, as well as the SOTA XBA design in \cite{rmntt} and the in-SRAM design in \cite{bpntt}. Our results show that \name can achieve significantly better throughput-per-area performance than both of these solutions, even as the parameter size increases. This highlights how \name  can lead to decreased area consumption of the XBA-based solution without compromising  the throughput. \eat{Our new data mapping technique reduces the peripheral cost and high memory density of the emerging ReRAM device.}

\eat{
\begin{figure}[b]
    \centering
    \includegraphics[width=\linewidth]{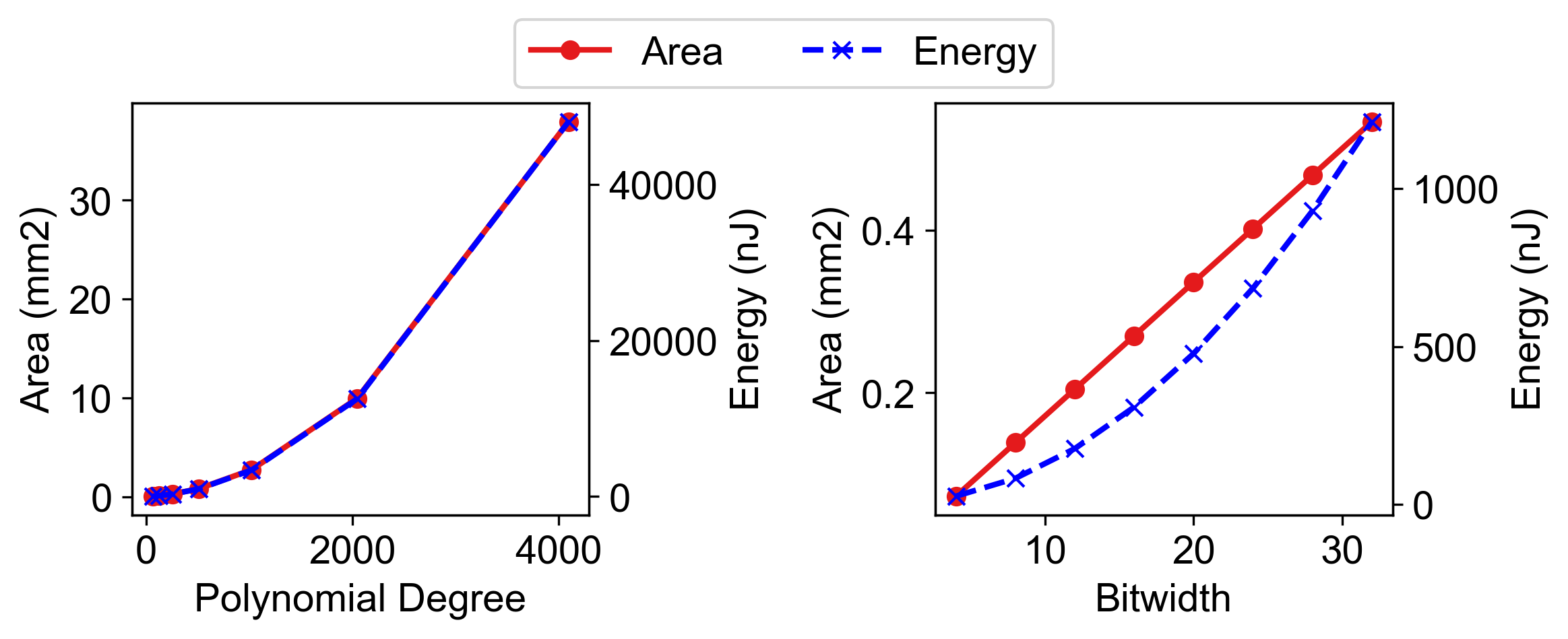}
    \caption{Impact of polynomial degree and bitwidth on area and energy of \name.}
    \label{fig:latency}
\end{figure}}

\begin{figure}
    \centering
    \includegraphics[width=\linewidth]{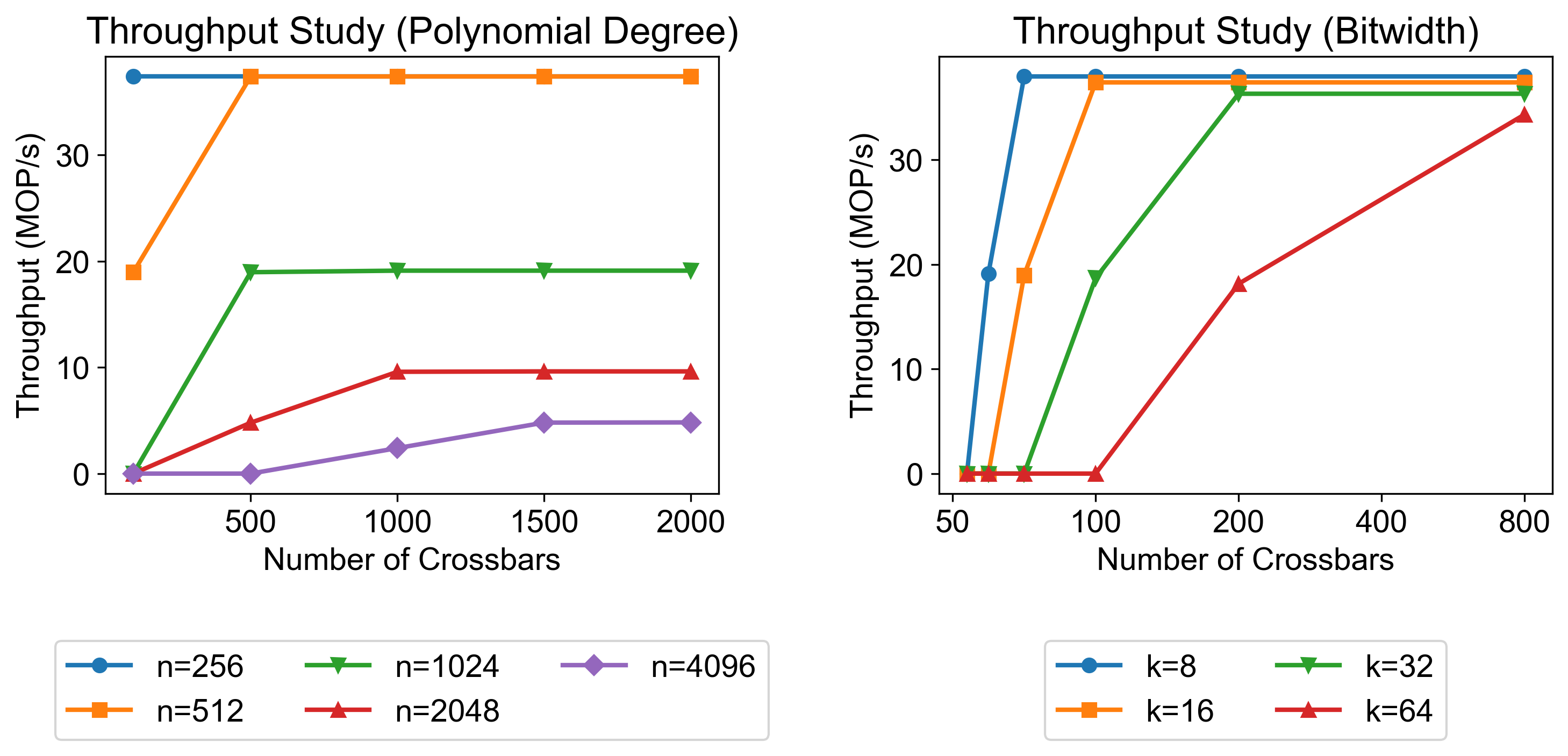}
    \caption{Scalibility study shows the throughput of \name under different polynomial degrees and bitwidths given a fixed total XBA number.}
    \label{fig:throughput}
\end{figure}

\subsection{Scalibility of \name}
\label{sec:memory_utilized}

Modern applications like HE in privacy-preserving machine learning often choose polynomials with a large degree and bitwidth\cite{HE_F1,GAZELLE}. Storing these entirely within XBAs demands a high number of arrays, leading to significant area usage and energy consumption. \eat{As shown in Fig.~\ref{fig:latency}, we observed an increasing trend in area and energy as we increased both polynomial degree and bitwidth in our design.}

Our polynomial mapping scheme (Sec~\ref{sec:poly_mapping}) allows us to reuse arrays. This enables us to employ a smaller number of XBAs to accommodate larger polynomials. However, reusing XBAs could potentially affect our design's latency. To address this, we conducted an experiment where given a fixed number of XBA arrays, we assessed the capability of \name to adapt to various polynomial degrees and bitwidths. The objective here was to determine how we could optimize  \name to maximize design throughput under different polynomial degrees and bitwidths constraints.

The left graph in Fig.~\ref{fig:throughput} depicts the maximum throughput of \name using different numbers of XBAs. We considered polynomial degrees ranging from 256 to 2048, with a fixed bitwidth of 16. The right graph demonstrates the maximum throughput for different bitwidths ranging from 8 to 64, while maintaining a constant polynomial degree of 512. Our experiments highlight that our design is capable of managing a wide array of polynomial degrees and bitwidths while maintaining a fixed number of XBAs. As anticipated, higher degrees and bitwidths require longer computation times due to the necessity for reuse of the same arrays within the pipeline. By modifying the number of XBAs, we can manage the balance between area and throughput. Overall, our design showcases robust scalability, effectively adapting to a broad spectrum of polynomial degrees and bitwidths.

\section{Conclusion}
\label{sec:conclusion}
In summary, this paper proposes a novel PMM accelerator based on XBA-type CIM for accelerating the most time-consuming part of lattice-based cryptography algorithms. The proposed X-Poly design achieves 3.1 MOP/s throughput and offers 200$\times$ latency improvement compared to CPU-based implementations. It also achieves 3.9$\times$ throughput per area improvements compared with the SOTA CIM accelerators. The suitability of NTT-based solutions for CIM-based PMM acceleration is evaluated, and a novel bit mapping technique is proposed to reduce area and energy overhead. PE-level optimization is conducted to increase memory utilization and support different scales of problems with a fixed number of XBAs. 
\pagebreak

\bibliographystyle{IEEEtran}      
\bibliography{references}      

\end{document}